\documentclass[10pt]{article}

\usepackage{graphicx}
\graphicspath{{./figures/}}
\usepackage{rotating}

\usepackage{amsmath}
\usepackage{amsfonts}
\usepackage{amssymb}

\usepackage{url}

\begin{document}

\author{
Michael Menzel\\
Research Center for Information Technology\\
Karlsruhe Institute of Technology\\
Karlsruhe, Germany\\
\url{menzel@fzi.de} 
\and 
Rajiv Ranjan\\
School of CSE, University of New South Wales\\
Sydney, Australia\\
Information Engineering Lab, CSIRO ICT Center\\
Canberra, Australia\\
\url{rajiv.ranjan@csiro.au}
}

\title{
\begin{Large}
Technical Report
\end{Large}\\
CloudGenius: Automated Decision Support for Migrating Multi-Component Enterprise Applications to Clouds}

\maketitle

\begin{abstract}
One of the key problems in migrating multi-component enterprise applications to Clouds is selecting the best mix of VM images and Cloud infrastructure services. A migration process has to ensure that Quality of Service (QoS) requirements are met, while satisfying conflicting selection criteria, e.g. throughput and cost. When selecting Cloud services, application engineers must consider heterogeneous sets of criteria and complex dependencies across multiple layers impossible to resolve manually. To overcome this challenge, we present the generic recommender framework CloudGenius and an implementation that leverage well known multi-criteria decision making technique Analytic Hierarchy Process to automate the selection process based on a model, factors, and QoS requirements related to enterprise applications. In particular, we introduce a structured migration process for multi-component enterprise applications, clearly identify the most important criteria relevant to the selection problem and present a multi-criteria-based selection algorithm. Experiments with the software prototype CumulusGenius show time complexities.
\end{abstract}

\section{Introduction}

The emergence of Cloud computing \cite{armbrust2009above} over the past five years is potentially one of the breakthrough advances in the history of computing. Cloud computing paradigm is shifting computing from physical hardware- and locally managed software-enabled platforms to virtualized Cloud-hosted services. Cloud computing assembles large networks of virtualized services: hardware services (compute services, storage, and network) and infrastructure services (e.g., web server, databases, message queuing systems, monitoring systems, etc.). Cloud providers including Amazon Web Services (AWS), Microsoft Azure, Rackspace, GoGrid, and others give users the option to deploy their application over a pool of virtually infinite resources with practically no capital investment and with modest operating cost proportional to the actual use. Elasticity, cost benefits, and abundance of resources motivate many organizations to migrate their enterprise applications to Clouds.  

There are two application engineering layers for compute services: a) software resource, where an engineer builds applications using APIs provided by the Cloud. A software resource (also referred to as an appliance or VM image) is a pre-configured, virtualization-enabled, self-contained, and pre-built Virtual Machine (VM) image  that can be integrated with other compatible VM images for architecting complex applications. Major providers at this layer include thecloudmarket.com, 3Tera Applogic, and BitNami; and b) Infrastructure as a Service (IaaS) (hardware resources), where an engineer runs software applications on compute services, using the APIs provided to leverage other infrastructure services. A VM instance is essentially a piece of virtualization software (e.g. Xen, KVM, etc.) running on physical Cloud servers. It is the most common method of exposing the computational power (e.g. CPU cores, physical memory, storage capacity, etc.) to software applications. Amazon EC2, GoGrid, and Rackspace are among the major providers of virtualized hardware resources as services.

\begin{figure}[t]
\centering
\includegraphics[width=0.5\columnwidth]{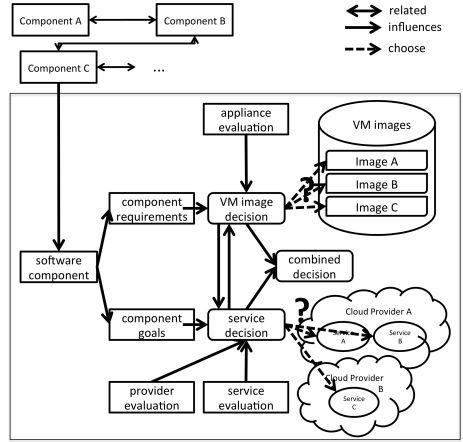} 
\caption{Overview of the Selection Problem}\label{overview-selection}
\end{figure}

\textbf{Problem Statement.} Enterprise applications (e.g. customer relationship management, employee payroll, and supply chain management) can typically be decomposed into three software resource layers: i) front-end web servers to handle end-user requests and application presentation; ii) business logic to perform specialized application logic; and iii) back-end database servers. The flow of requests between these layers is often complex. Each layer may instantiate multiple software resources; each software resource may need to be replicated on multiple compute resources, while load-balancers distribute requests across each instance of VM images. This creates an enterprise application consisting of multiple components: an IT system formation. Optimal application QoS demands bespoke configuration both at software and IaaS layer, yet no detailed, comprehensive cost, performance or feature comparison of Cloud services exists. The key problem in mapping applications in form of multi-component IT system formations to Cloud resources is selecting the best size and mix of software and hardware resources to ensure that application QoS targets are met, while satisfying conflicting selection criteria \cite{SPE:SPE1110} related to software (e.g. virtualization format, operating system, etc.) and hardware (e.g. maximizing throughput, minimizing cost) resources. For instance, before mapping a Bitnami's Web server appliance \cite{bitnami2011} to a Amazon EC2\cite{awsec22011} virtual machine instance resource, one needs to consider whether they are compatible in terms of virtualization format (e.g., VMWare, AMI, etc.) and other system-level constraints (e.g., Unix or Windows operating system).
Figure \ref{overview-selection} depicts the selection problem of migrating multi-component enterprise applications, IT system formations, to Cloud infrastructure services.

\begin{figure}
\centering
\includegraphics[width=0.5\columnwidth]{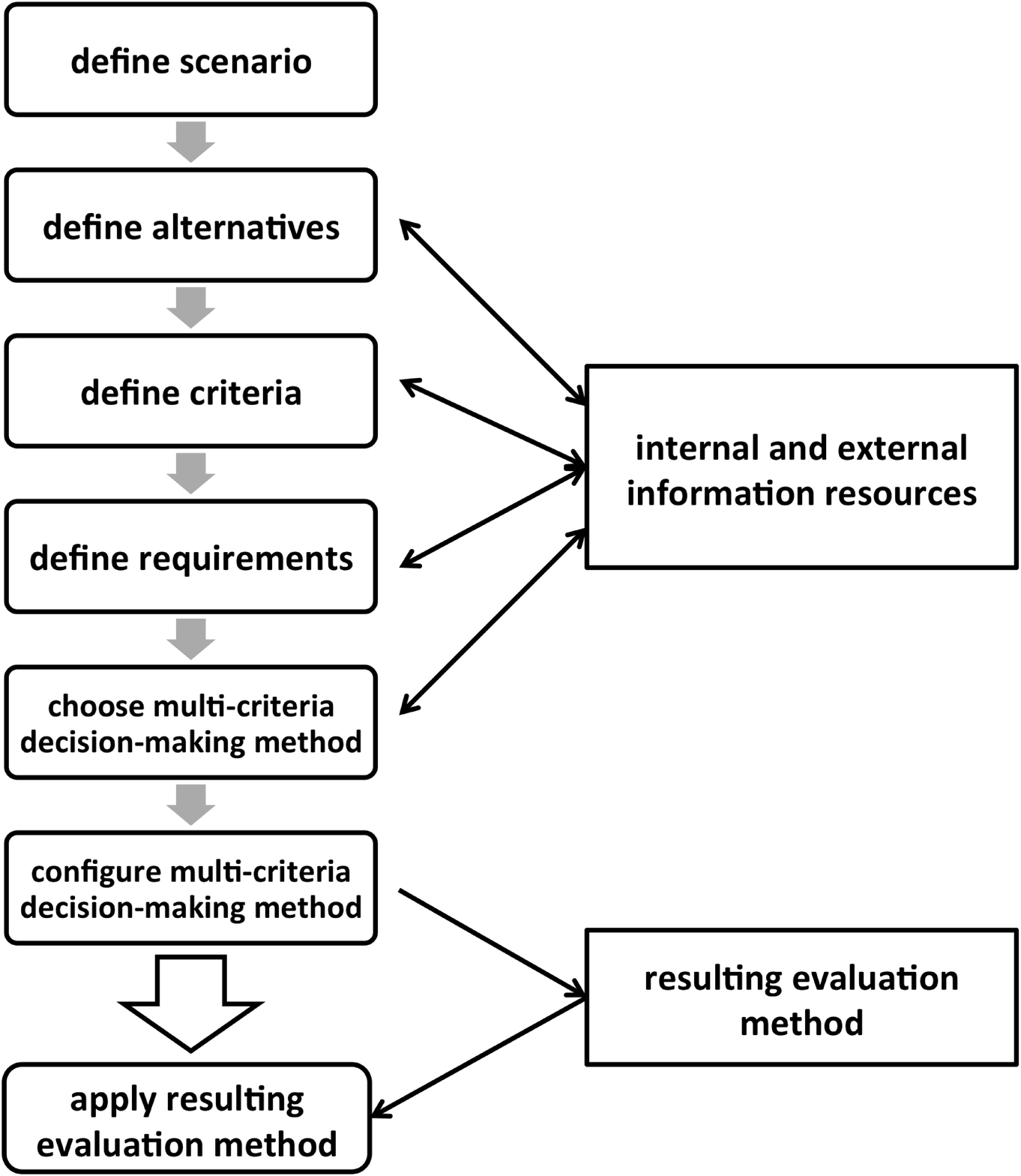} 
\caption{Overview of the $(MC^2)^2$ Process}\label{mc22-process}
\end{figure}

\textbf{Proposed Approach.} To counter the above complexities for migrating multi-component enterprise applications, we propose a novel, flexible, and automated decision-making framework, CloudGenius, that translates Cloud service selection steps into multi-criteria decision-making problems to determine the most applicable VM images (at software resource layer) and compatible compute services (at IaaS layer). CloudGenius provides a framework that guides through a Cloud migration process and offers a model and methods to determine the best combined and compatible choice of VM images and services. The framework leverages the evaluation and decision-making framework $(MC^2)^2$ \cite{menzel2010} for supporting  multi-criteria-based selection. The $(MC^2)^2$ framework provides a process depicted in Figure \ref{mc22-process} allowing the creation of an evaluation method that comprises a requirements check and evaluates multiple alternatives on an absolute $(0-1)$ scale. Within $(MC^2)^2$ process our approach proposes the Analytic Hierarchy Process (AHP) as the multi-criteria decision-making method of choice. AHP allows for administrator trade-offs and compensation between multiple criteria and, hence, give alternatives with a low value regarding one criterion the chance to retain within the set of feasible solutions. Compensations are influenced by the weighting which is derived from the trade-offs users make in pair-wise comparisons of criteria. 

The paper is structured as follows. First, we reflect related work in Section \ref{related} and, afterwards, we present the CloudGenius framework in Section \ref{framework}. The framework introduces a migration process and formal model which forms the basis for decision support within the process. Further, we present CumulusGenius, a prototypical implementation of the framework in Section \ref{prototype}. In Section \ref{experiments} we present the results of experiments on the time complexity of CumulusGenius before we discuss future work and conclude in Section \ref{discussion-future}.

\section{Related Work}\label{related}

Multiple approaches have been introduced by the Web service community that define multi-component Web services \cite{hamadi2003petri}\cite{zhang2008colored} but do not address the characteristics of the Cloud.
Existing work in the Cloud context provides provider or service evaluation methods but lacks multi-component support\cite{chan2010ranking}.
Multiple approaches for multi-component setups in the Cloud have applied optimization \cite{ye2011genetic}\cite{SPE:SPE1032}\cite{Goudarzi_Pedram}\cite{hajjat2010cloudward} and performance measurement techniques \cite{li2010cloudcmp} for selecting hardware resources (provider side) or Cloud infrastructure services (client side) according to quantitative criteria (throughput, availability, cost, reputation, etc.). While doing so, they have largely ignored the need for VM images, a migration process with transparent decision support and adaptability to custom criteria, and, hence, lack flexibility.

Additionally, there is preliminary work that provides decision support for selecting VM images and infrastructure services. Dastjerdi et al. \cite{dastjerdi2010effective} propose an approach that selects Cloud VM images and Cloud infrastructure services with an ontology-based requirements check but lacks a service evaluation.
Khajeh-Hosseini et al. \cite{khajeh2010cloud} \cite{khajeh2011decision} developed the Cloud Adoption Toolkit that offers a high-level decision support for IT system migration of enterprises. The focus of the decision support is on risk management and a cost model which incorporates expected workload on the IT system.

\section{CloudGenius Framework}\label{framework}

A migration of an IT system formation to Cloud infrastructures is complex and demands the choice of adequate Cloud infrastructure services and Cloud VM images for every component within the formation. We propose CloudGenius, a framework that guides through a Cloud migration process that provides methods that support multi-criteria-based decisions on selecting a Cloud VM images and Cloud infrastructure services component-wise. 
In the following subsections we present the process and give details on the formal model of the selection problem, the required user input and flexibilities, and the selection and combination steps that choose an image and service from the abundance of offerings and find the best combination. Finally, an alternative evaluation variant is addressed.



\subsection{Multi-Component Cloud Migration Process}

\begin{sidewaysfigure}[!p]
\centering
\includegraphics[width=\textheight]{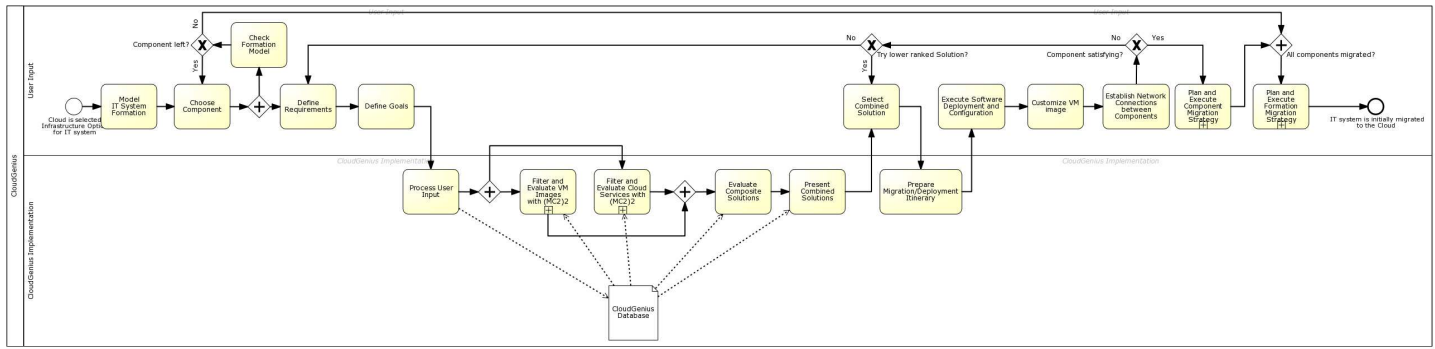} 
\caption{Multi-Component Migration Process of the CloudGenius Framework}\label{cloudgenius-migration-process}
\end{sidewaysfigure}

Figure \ref{cloudgenius-migration-process} depicts CloudGenius' migration process for IT system formations in Business Process Model and Notation (BPMN) 2.0. The process is divided in two lanes: (1)  "user input" lane with domain experts such as application engineers providing input and (2) "CloudGenius" lane where steps are completed by an implementation of the framework. 
The process allows for a loop enabling a component-wise migration and cycles for step-wise, incremental improvements of every component's migration. Within the cycles engineers have to define requirements and preferences and CloudGenius applies the $(MC^2)^2$ decision-making framework to recommend a ranked VM image and Cloud service combinations for a certain component. 

\subsection{Formal Model of CloudGenius}

A formal mathematical model is introduced to formalize the problem addressed by the CloudGenius framework. The model defines all parameters involved in the problem which the evaluation is based on.
Table \ref{cloudgenius-math-model} summarizes parameters of the model. 

\begin{table}[!h]
\centering
\caption{CloudGenius' Formal Model}\label{cloudgenius-math-model}
\begin{tabular}{|l|p{4in}|} \hline
Parameter&Description\\ \hline
$l$&number of software components\\
$m$&number of Cloud VM images\\
$n$&number of Cloud infrastructure services\\
$o$&number of Cloud providers\\
$r$&number of formulated Requirements\\
$C$&set of software components\\
$I$&set of relations between components\\
$A$&set of Cloud VM images\\
$S$&set of Cloud infrastructure services\\
$P$&set of Cloud providers\\
$R_A$&set of Cloud image requirements\\
$R_S$&set of Cloud service requirements\\
$D$&set of image-service compatibilities\\
$E$&set of inter-image compatibilities\\
$F$&set of inter-service compatibilities\\
$\hat{A}_{a_i}$&set of numerical attributes of i-th image\\
$\hat{A}_{s_j}$&set of numerical attributes of j-th service\\
$\hat{B}_{a_i}$&set of non-numerical attributes of i-th image\\
$\hat{B}_{s_j}$&set of non-numerical attributes of j-th service\\
$c_h$&h-th software component to be migrated to the Cloud\\ 
$a_i$&i-th image in the set of m images\\
$s_j$&j-th service in the set of n services\\ 
$p_k$&k-th provider in the set of o providers\\
$d_l$&l-th dependency $(a_i, s_j)$ in the set $D$\\ 
$\tau$&Cloud image or service $\tau \in A\cup{}S$\\
$\chi(\alpha)$&Value of numerical attribute $\alpha$ in CloudGenius database\\ 
$\chi(\beta)$&Value of non-numerical attribute $\beta$ in CloudGenius database\\ 
$\alpha_{h,{\tau_i}}$&h-th numerical attribute of i-th $\tau$\\
$\beta_{g,{\tau_i}}$&g-th non-numerical attribute of i-th $\tau$\\
$v_{\tau_i}$&value of i-th $\tau$ calculated with $(MC^2)^2$\\
\hline
\end{tabular}
\end{table}

The model of CloudGenius consists of $l$ components which are part of a formation, $m$ images $a_i$, $n$ services $s_j$ and $o$ providers $p_k$. $C$ is the corresponding set of software components reflecting the formation, $A$ the set of VM images, $S$ the set of Cloud infrastructure services and $P$ the set of Cloud providers. $a_i$ and $s_j$ own numerical, measurable and non-numerical attributes of the sets $\hat{A}_{a_i}$, $\hat{A}_{s_j}$, $\hat{B}_{a_i}$ and $\hat{B}_{s_j}$. $\chi$ represents a value connected with a numerical attribute $\alpha$ or non-numerical attribute $\beta$. Furthermore, the model introduces $r$ requirements and a goal/criteria hierarchy including $g$ goals and $c$ leaf criteria. 

Based on the model CloudGenius determines the best combination ($a_i$, $s_j$) where $a_i$ and $s_j$ are the image and infrastructure service of provider $p_k$ that have the highest value of all combinations according to the user's preferences and with $a_i$ deployable on $s_j$. 

An evaluation method built with $(MC^2)^2$ can be employed to determine the best image $a_i \in A$ and service $s_j \in S$. Therefore, $(MC^2)^2$ is interpreted as $f(a_i, \hat{A}_{a_i}, \hat{B}_{a_i})\mapsto{}v_{a_i}$ which allows to find\\ $max \{v_{a_1}, ..., v_{a_m}\}$, and a function $f(s_j, \hat{A}_{s_j}, \hat{B}_{s_j})\mapsto{}v_{s_j}$ upon which $max \{v_{s_1}, ..., v_{s_n}\}$ can be determined. The results can be merged to the function $f(a_i,s_j)\mapsto{}v_{(a_i, s_j)}$ in order to determine a combined value.

Only feasible combinations of an image and service combination must be considered, meaning an image has to be deployable on a service. The feasibility of a combination is indicated by the set $D$ which holds all compatibility dependencies between images and services. The feasibility of a multi-component formation is defined by set $E$ defining VM image compatibilities and set $F$ defining service compatibilities. $E$ and $F$ hold pairs of VM images resp. Cloud services that are compatible.

\subsection{IT System Formation Model}

In an initial step, stakeholders of the migration process, typically application engineers, have to define an IT system formation in the model. All components of the formation setup must be added as component items $c$ to set $C$ and interconnections must be defined in form of component pairs  in set $I$. Further, a software feature must be assigned that categorizes the component. The set of available features comprises Web Server, Application Server, ERP system and CRM system and limits the set of plausible VM images. Next, the CloudGenius migration process supports engineers in the component-wise realization of the formation.

\subsection{Software Component Requirements \& Preferences}\label{sec-reqs-prefs}

Within CloudGenius' process, after a software component has been chosen, domain experts resp. engineers have to formulate requirements and preferences. Requirements formulation comprises choosing an attribute to set a constraint on and specifying a minimum or maximum value for numerical values, and a set of allowed items for non-numerical values. Table \ref{requirement-types} gives an overview of four requirement types aligned with the $(MC^2)^2$ framework which uses con-/disjunctive satisficing methods. The table assumes $\chi$ to be the attribute value under consideration, $v_r$ to be the given numerical requirement value, $s$ a given non-numerical value and $S$ a given set of non-numerical values. The applied requirements check are formulated in boolean expressions.
Attributes available for requirements definitions can be drawn from attributes of Cloud VM images and Cloud infrastructure services listed in Table \ref{image-numerical-attributes}, \ref{image-nonnumerical-attributes}, \ref{service-numerical-attributes} and \ref{service-nonnumerical-attributes}. Requirements set upon attributes of VM images belong to set $R_A$, requirements for services to $R_S$.

\begin{table}
\centering
\caption{CloudGenius Requirement Types}\label{requirement-types}
\begin{tabular}{|c|c|l|} \hline
Value Type&Req. Type&Boolean\\ \hline
Numerical&Max&$\chi(\alpha) < v_r$\\
Numerical&Min&$\chi(\alpha) > v_r$\\
Non-numerical&Equals&$\chi(\beta) = s$\\
Non-numerical&OneOf&$\chi(\beta) \in S$\\
\hline\end{tabular}
\end{table}

\begin{figure}[h]
\centering
\includegraphics[width=0.5\columnwidth]{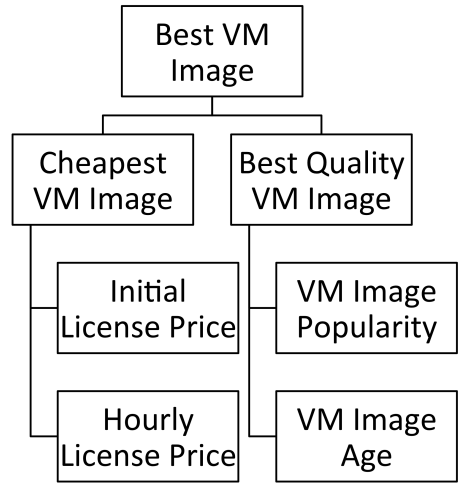} 
\caption{CloudGenius VM Image Goal Hierarchy}\label{overview-image-goals}
\end{figure}

\begin{figure}[h]
\centering
\includegraphics[width=0.7\columnwidth]{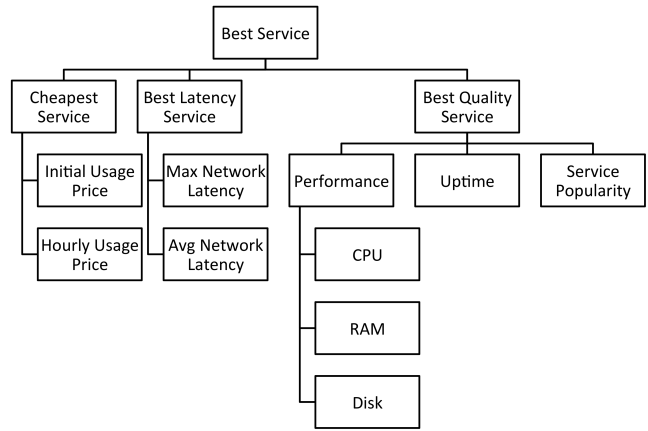} 
\caption{CloudGenius Infrastructure Service Hierarchy}\label{overview-service-goals}
\end{figure}

Stating preferences is carried out by selecting and weighting given goals and criteria. CloudGenius proposes hierarchies of goals and criteria and asks engineers to select and weight the items in pair-wise comparisons (in analogy to AHP). Figure \ref{overview-image-goals} and Figure \ref{overview-service-goals} depict the proposed hierarchies of goals and criteria offered by CloudGenius. Goals are high level and group critera while criteria are associated with attributes to be evaluated.  

Since the $(MC^2)^2$ evaluation framework would support more complex goal hierarchies CloudGenius is not limited in this aspect. However, a higher complexity of the goal hierarchy increases the amount of input a user must provide and opposes CloudGenius' goal of being lightweight. Possible additional goals are QoS proposed by Kalepu et al. \cite{kalepu2003verity} such as reliability, accessibility or support.

Finally, domain experts have to state the importance of a Cloud VM image's and a Cloud infrastructure service's value within a combined value in a pair-wise comparison.

\subsection{Cloud VM Image Selection}

VM images can be characterized by functional and non-functional attributes with numerical or non-numerical values assigned stating technical details or qualities. Table \ref{image-numerical-attributes} gives an overview of the measurable, numerical attributes the CloudGenius framework proposes. The table lists their influence direction, metrics and the range of possible values. A list of non-numerical attributes is presented in Table \ref{image-nonnumerical-attributes}. The variability of an attribute represent the ability to change the value over life time of a VM image, e.g. the popularity of an image might dynamically change over time and its operating system is static. In Table \ref{image-nonnumerical-attributes} all non-numerical attributes are static. Numerical attributes, additionally, have either a negative or positive influence on the value the higher their value. For instance, the higher the license costs the less interesting becomes an image.

\begin{table}[h]
\centering
\caption{Cloud Image Numerical Attributes}\label{image-numerical-attributes}
\begin{tabular}{|l|l|l|l|l|} \hline
Name&Influence&Variability&Metric&Value Range\\ \hline
Hourly License Price&Negative&Dynamic&\$/h&0-$\infty$ \$/h\\
Popularity&Positive&Dynamic&\%&0-100 \%\\
Age&Positive&Dynamic&Days&0-$\infty$ Days\\
OS Version&None&Static&Version&0-$\infty$\\
Software Version&None&Static&Version&0-$\infty$\\
\hline\end{tabular}
\end{table}

\begin{table}[h]
\centering
\caption{Cloud image Non-numerical Attributes}\label{image-nonnumerical-attributes}
\begin{tabular}{|l|l|} \hline
Name&Example Values\\ \hline
Virtualization Format&Xen, VMWare, \ldots\\
Operating System (OS)&Linux, Windows, \ldots\\
Implementation Language&Java, Perl, Ruby, \ldots\\
Software Feature&Web Server, Application Server, \ldots\\
Software&Apache HTTP, JBoss Appl. Server, \ldots\\
\hline\end{tabular}
\end{table}

The selection of attributes is drawn from own observations and literature on VM images and services \cite{dastjerdi2010effective}\cite{kalepu2003verity}, providing a basic set of attributes essential to Cloud VM image selection. Service attributes have a limited applicability to VM images being only the software resource. Therefore, we hope to be able to build a list of important attributes from usage data of a publicly available prototype\cite{cumulusgenius2011,cumulusgeniusonline2011} and VM image databases such as thecloudmarket.com\cite{cloudmarket2011} or BitNami\cite{bitnami2011}.

The search for a proper VM image is hugely influenced by the categorization of a VM image, the software feature, that must match a components feature. CloudGenius supports generally all sorts of software features but we limit our focus on enterprise application systems and provide general attributes for those types of systems. Support for more or more detailed features for specific enterprise application systems requires feature-specific attributes.

Evaluation of VM images requires an evaluation method to be built with $(MC^2)^2$ based on domain experts' input and CloudGenius' formal model. 
In $(MC^2)^2$ all images from a database which holds the set of images including all image attributes become the evaluation alternatives.
Further, all influential (influence $\neq$ none) numerical attributes of VM images (see Table \ref{image-numerical-attributes}) become leaf criteria in the customized goal and criteria hierarchy weighted earlier by engineers (see Section \ref{sec-reqs-prefs}).
CloudGenius defines all requirements set on image attributes and defines AHP to be the multi-criteria decision-making method.

With $(MC^2)^2$ an evaluation method is created that matches a function $f(a_i, \hat{A}_{a_i}, \hat{B}_{a_i})\mapsto{}v_{a_i}$ which is applied on all images $a_i$. By setting all $\chi(\alpha)$ as parameters in the function a value $v_{a_i}$ and image $a_i$ can be determined with $v_{a_i} = 0$ whenever requirements are violated. In detail, $(MC^2)^2$ creates the function described in Equation \ref{function-image-evaluation} where $w_j$ is the global criteria weight of j-th positive and normalized numerical attribute with value $\chi(\alpha_{j,a_i,+}$. $f(a_i, \hat{A}_{a_i}, \hat{B}_{a_i})$ reflects the multiplicative index of AHP that divides positive goals by negative ones.
\begin{equation}\label{function-image-evaluation}
f(a_i, \hat{A}_{a_i}, \hat{B}_{a_i}) = 
\begin{cases}
\sum\limits_{j=0}^{|\hat{A}_{a_i}|} w_j\chi(\alpha_{j,a_i,+})\over{\sum\limits_{j=0}^{|\hat{A}_{a_i}|} w_j\chi(\alpha_{j,a_i,-})} & \forall r \in R_A: r=true\\
0 & \text{else}
\end{cases}
\mapsto{} v_{a_i}
\end{equation}

In case, none of the images meets all requirements in a subsequent step all images that meet all but one requirement in $R_A$ are considered. This procedure repeats until a non-empty set of images is found that fulfills a smaller number of requirements.

To find the best image $max \{v_{a_1},...,v_{a_m}\}$ is to rank all images by the resulting values comparable on an absolute [0, 1] scale.

\subsection{Cloud Infrastructure Service Selection}

In parallel to a VM image selection, for the selection of a Cloud compute service, such as Amazon EC2 or Joyent Public Cloud, CloudGenius proposes a range of numerical and non-numerical attributes and leverages the $(MC^2)^2$ framework to gain an evaluation method $g(s_j, \hat{A}_{s_j}, \hat{B}_{s_j})\mapsto{}v_{s_j}$ formulated in Equation \ref{function-service-evaluation}.

\begin{equation}\label{function-service-evaluation}
g(s_j, \hat{A}_{s_j}, \hat{B}_{s_j}) = 
\begin{cases}
\sum\limits_{i=0}^{|\hat{A}_{s_j}|} w_i\chi(\alpha_{i,s_j,+})\over{\sum\limits_{i=0}^{|\hat{A}_{s_j}|} w_i\chi(\alpha_{i,s_j,+})} & \forall r \in R_S: r=true\\
0 & \text{else}
\end{cases}
\mapsto{} v_{s_j}
\end{equation}

A list of numerical attributes for the Cloud service selection evaluation method is given in Table \ref{service-numerical-attributes} (all dynamic), non-numerical attributes are given in Table \ref{service-nonnumerical-attributes} (all static).

\begin{table}[!h]
\centering
\caption{Cloud Service Numerical Attributes}\label{service-numerical-attributes}
\begin{tabular}{|l|l|l|l|} \hline
Name&Influence&Metric&Value Range\\ \hline
Hourly CPU Price&Negative&\$/h&0-$\infty$ \$/h\\
Network Send Price&Negative&\$/B&0-$\infty$ \$/Byte\\
Network Recieve Price&Negative&\$/B&0-$\infty$ \$/Byte\\
Internet Send Price&Negative&\$/h&0-$\infty$ \$/h\\
Internet Recieve Price&Negative&\$/B&0-$\infty$ \$/Byte\\
CPU Perfomance&Positive&Flops&0-$\infty$ Flops\\
CPU Cores&Positive&Cores&0-$\infty$ Cores\\
RAM Perfomance&Positive&Flops&0-$\infty$ Flops\\
RAM Size&Positive&Bit&0-$\infty$ Bit\\
Disk Perfomance&Positive&Flops&0-$\infty$ Flops\\
Disk Size&Positive&Bit&0-$\infty$ Bit\\
Max. Latency&Negative&ms&0-$\infty$ ms\\
Avg. Latency&Negative&ms&0-$\infty$ ms\\
Uptime&Positive&\%&0-100 \%\\
Service Popularity&Positive&\%&0-100 \%\\
\hline\end{tabular}
\end{table}

\begin{table}[!h]
\centering
\caption{Cloud Service Non-numerical Attributes}\label{service-nonnumerical-attributes}
\begin{tabular}{|l|l|} \hline
Name&Example Values\\ \hline
Provider&Amazon, Rackspace, \ldots\\
Location Country&Germany, Australia, \ldots\\
\hline\end{tabular}
\end{table}

Cloud infrastructure services own multiple numerical attributes that imply a measurement or benchmarking. Hourly Usage, Network and Internet Traffic Prices are typically provided from the provider or must be calculated with a provider's price model or cost calculation schemes \cite{klems2009clouds}\cite{khajeh2011decision}. For performance and latency attributes benchmarking tools are required as those are often not provided by Cloud providers \cite{lenk2011you} in contrast to attributes such as CPU core or RAM size. The uptime of a service is a long-term experience that can be provided as guaranteed uptime by a provider or his SLA, or from user experiences. In parallel, service popularity can be gained from user experiences.

In parallel to VM image selection an evaluation method is created with $(MC^2)^2$. With all parameters set (VM images as alternatives, a criteria hierarchy built from attributes, requirements $R_S$ and criteria weights) the $(MC^2)^2$ process can be completed and a new Cloud service evaluation method based on AHP is created. The Cloud service evaluations can be retrieved from applying the new evaluation method $g(s_j, \hat{A}_{s_j}, \hat{B}_{s_j})$. The highest ranked Cloud infrastructure alternative is $max \{v_{s_1}, ..., v_{s_n}\}$.

\subsection{Best Combination}

In the final evaluation step, VM images and Cloud services are combined to solutions. Every VM image $a_i$ is combined with a service $s_j$ to create a possible solution pair $(a_i, s_j)$. The CloudGenius model holds the set $D$ of all compatibility dependencies between images and services where an existing dependency between image $a_i$ with service $s_j$ implies feasibility and, hence, a value $v_{a_i,s_j} > 0$. Equation \ref{function-composite-evaluation} formulates the function that maps solution pairs to a value using an operator $\bullet$. For more complex combination value computations a function $h(f(\cdot),g(\cdot))$ may return the overall value based on the image and service evaluation functions $f$ and $g$ instead of an operator $\bullet$ only. 

\begin{equation}\label{function-composite-evaluation}
f(a_i,s_j) = \\
\begin{cases}
f(a_i,\hat{A}_{a_i},\hat{B}_{a_i}) \bullet g(s_j,\hat{A}_{s_j},\hat{B}_{s_j}) & (a_i,s_j) \in D\\
0 & else\\
\end{cases}
\mapsto{} v_{a_i,s_j}
\end{equation}

CloudGenius promotes the $+$ operator to sum up VM image and service value to a total value of a combination. The sum of both values allows for compensation between image and service selection where a low quality service and high quality image might have a lower total value than a combination of medium quality image and service. In contrast, $*$ or $\times$ operator helps finding VM image infrastructure service combinations with most balanced values. Equation \ref{function-composite-evaluation-weights} gives the evaluation function with user defined weights $w_a$ and $w_s$ that sum up to 1 that describe the importance of the VM image and service within a combined solution.
Services of a Cloud provider that are located at different locations might cause internet traffic costs billed by the provider. 
Equation \ref{function-network-traffic-delta} shows the $\Delta$ of total network traffic costs (internet and local network) for a component $i \in C$ connected with other components according to $I$ where $I_i$ holds $i$'s relations. The $\Delta$ represents the extra costs for network traffic introduced by the combined solution added as component to the multi-component formation. Let $c_{i,o,R_l}$, $c_{i,o,S_l}$, $c_{i,o,R_g}$, $c_{i,o,S_g}$ be expected cost of incoming and outgoing local network and internet traffic between components $i$ and $o$, given by domain experts. All $\Delta_{\text{network,i}}$ should be normalized to $(0, 1)$ scale, e.g. using AHP with one criterion, to follow AHP multiplicative index propositions.

\begin{equation}\label{function-composite-evaluation-weights}
f(a_i,s_j) = \\
\begin{cases}
w_a f(a_i,\hat{A}_{a_i},\hat{B}_{a_i}) + w_s g(s_j,\hat{A}_{s_j},\hat{B}_{s_j})\over{\Delta_{\text{network},c_h,\text{normalized}}} & (a_i,s_j) \in D, c_h \in C\\
0 & \text{else}\\
\end{cases}
\mapsto{} v_{a_i,s_j}
\end{equation}

\begin{equation}\label{function-network-traffic-delta}
\Delta_{\text{network,i}} = \sum^{I_i}_{o}\\
\begin{cases}
c_{i,o,R_l} + c_{i,o,S_l} & \text{provider and location equal}\\
c_{i,o,R_g} + c_{i,o,S_g} & \text{else}\\
\end{cases}
\end{equation}

The best overall solution pair to be added to the IT system formation has a value $max \{v_{a_1,s_1},...,v_{a_1,s_1}\}$. In a multi-component setup, however, relations between component restrain the number of actually feasible combined solutions. Thus, introduced constraints are defined by sets $E$ and $F$ allowing only multi-component setups where compatible VM images and services are composited. An added combined solution must be tested for its co-operation compatibility with a formation's components it has relations with according to set $I$.

After automated decision support, the migration process continues with a selected combined solution, deployment of the VM image on the service and further customization and re-evaluation cycles by engineers. With all components deployed and customized an applied migration strategy results in the multi-component IT system available on Cloud infrastructure services.

\subsection{Integrated Evaluation Approach}

\begin{figure}[h]
\centering
\includegraphics[width=0.55\columnwidth]{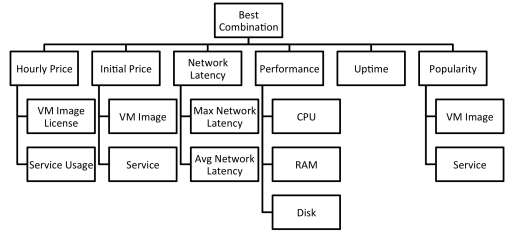} 
\caption{CloudGenius Integrated Goal Hierarchy}\label{overview-goals-integrated}
\end{figure}

Alternatively, a best combination can be determined in a single $(MC^2)^2$ process that process uses a single criteria hierarchy depicted in Figure \ref{overview-goals-integrated}. 

In this variant the compensatory relation between VM image and service are influenced by weighting of the criteria hierarchy. Due to the lack of transparency regarding partial results we do not promote this variant and leave it to an implementation to provide any or both evaluation options to a user.

\subsection{Complexity of CloudGenius Approach}

The whole problem addressed by CloudGenius seems rather complex and involves a number of calculations. Hence, analyzing the actual computational complexity of the approach becomes of interest to ensure its applicability. To our best knowledge we define $O$ of CloudGenius as following:
\begin{align*}
O(l * (
&\underbrace{m*|\hat{B_a}|+n*|\hat{B_s}|}_{\text{requirements check}} +\\
&\underbrace{m*|\hat{A_a}|+n*|\hat{A_s}|}_{\text{evaluation}} +\\
&\underbrace{m*n*|D|}_{\text{feasibility check}} + \underbrace{m*n}_{\text{combined value}}) + \underbrace{2* {(l-1)*(l-2)}\over{2}}_{\text{network costs send, recieve}}
)
\end{align*}

The computational complexity is mainly influenced by the number of VM images $m$, the number of services $n$ and the number of attributes resp. criteria. Part of the computations are requirements checks for all $m$ images and $n$ service regarding all requirements in $\hat{B_a}$ and $\hat{B_s}$. The evaluation adds computation steps for the VM image and service evaluation which depend on the number of images and services and the number of attributes. The defined $O$ only counts the number of evaluation steps to sum up a total value for an image or service. In more detail, AHP introduces more steps to normalize matrices and derive global weights. For simplicity these steps are omitted. Computations required for combined solutions contain the filtering of infeasible solutions by making all possible $m*n$ combinations and comparing with $D$. Finally, a combined value for all combinations is calculated. 
The complexity must be multiplied by the number of components $l$ and network costs are added in a multi-component setup. The summed effort for all network cost calculations is doubled to reflect send and recieve operations. The calculation effort is $(l-1)*(l-2)\over{2}$ as there is no network costs calculation effort for the first step where a first migrated component is added to the formation.

\section{CumulusGenius: An Implementation}\label{prototype}

With CumulusGenius \cite{cumulusgenius2011} we present an implementation of the decision-making support of the framework. The CumulusGenius java library offers a data model that enables the evaluation of VM images, Cloud infrastructure services and best combinations programmatically. A Web frontend that supports the framework's process and provides a database of VM images and Cloud services of the current Cloud provider landscape is currently under development \cite{cumulusgeniusonline2011}.

\section{Experiments}\label{experiments}

\begin{figure}[h]
\centering
\includegraphics[width=0.7\textwidth]{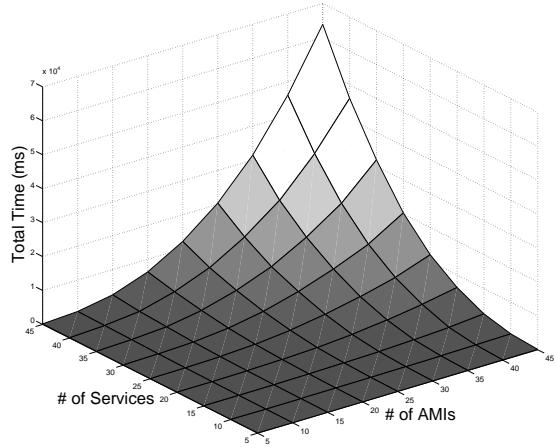} 
\caption{Total Time for Variable VM Images, Services (3 Components, 3 Providers)}\label{experiments_total}
\end{figure}

\begin{figure}[h]
\centering
\includegraphics[width=0.7\textwidth]{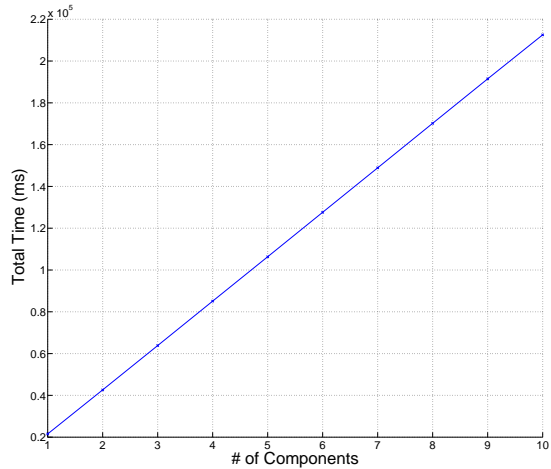} 
\caption{Total Time Per Component (45 VM Images, 45 Services, 3 Providers)}\label{experiments_incr_comp}
\end{figure}

We tested our implementation CumulusGenius in experiments on a test machine with Intel Core i7 2.7 Ghz and 8 GB of RAM. The experiments allow analyzing the resulting time complexity of CumulusGenius. The parameters of the experiments are the number of VM images, service and components. VM images and services are synthetically generated with all attributes having random values. There is a fix number of three providers and no requirements are defined to keep a full search space of combined solutions. Components are randomly assigned to a provider and all inter-connected to one another. When components are assigned to the same provider low network costs occur, in case of different providers high internet costs are assumed. Figure \ref{experiments_total} depicts the exponentially growing total time to find a solution for three components and Figure \ref{experiments_incr_comp} the linearly growing computation time when adding components. Evaluation of combined solutions produces the major part of the computation time.

\section{Conclusion \& Future Work}\label{discussion-future}

In this paper, we presented the CloudGenius recommender framework that transforms Cloud service selection from manual time-consuming scripting to a process that is flexible, and to a large extend automated. It provides a migration process and helps enterprise application engineers to select best resource mix at both software and IaaS layers over provider boundaries. We believe that CloudGenius framework leaves space for a range of enhancements and, yet, provides an amicable approach. To our knowledge no existing approach has addressed the problem of inter-dependencies between software and IaaS layers while selecting software and hardware resources for Cloud-based engineering of enterprise applications.  

A major issue in Cloud service selection is the domain of the search space (i.e. the completeness of VM images and Cloud services database), the criteria catalogs, and the quality and correctness of measured values. To address these issues, we will focus on integrating existing benchmarking services such as CloudHarmony \cite{cloudharmony2011} in the CloudGenius framework. Work related to automated benchmarking is already in progress \cite{haak2011autonomic}. A critical mass of data on VM images and IaaS level services might be gained by integrating existing databases such as thecloudmarket.com \cite{cloudmarket2011} or CloudHarmony \cite{cloudharmony2011}. Further, we aim at making data decision and user-specific, like e.g. latency measurements. 

The proposed framework expects VM images to provide one feature instead of whole software stacks representing a whole formation or basic VM images only containing an OS. Future work should predict expected customization efforts and consider it in decisions. Additionally, explicit support for hybrid setups, quality concerns, in particular, reliability (choice of multiple, different VM images per component and geographical locations), and middleware and persistence layer services is future work. Also, CloudGenius' step-by-step approach depends on the component order within a migration what shall be overcome by global optimization in the future.  Moreover, requirement checks might be extended with a feature model-based approach \cite{wittern2011use}. Regarding automated application deployment we want to provide a process based language that considers control and data flow dependencies and elastic behavior of individual appliances.

It is planned to drive an evaluation with industry partners of the CumulusGenius prototype providing a Web-frontend \cite{cumulusgeniusonline2011}.

\section{Acknowledgments}
Initial research work on Cloud virtual machine image and infrastructure service data models was done when Dr. Rajiv Ranjan was employed at University of New South Wales on strategic eResearch grant scheme.

\bibliographystyle{abbrv}
\bibliography{references}

\end{document}